\documentclass{article}

\usepackage{amssymb,amsfonts,amsmath,stmaryrd}
\usepackage{cite,enumerate,float,indentfirst}
\usepackage{color}

\def\be{\begin{eqnarray}}
\def\ee{\end{eqnarray}}
\def\nn{\nonumber}

\def\p{\partial}

\def\Tr{{\rm Tr}\,}

\newcommand{\beq}{\begin{equation}}
\newcommand{\eeq}{\end{equation}}
\newcommand{\beqa}{\begin{eqnarray}}
\newcommand{\eeqa}{\end{eqnarray}}

\newcommand{\dP}{\zeta}

\definecolor{red}{rgb}{1,0,0}
\definecolor{orange}{rgb}{1,0.5,0}
\definecolor{violet}{rgb}{0.7,0,1}

\voffset= - 1.1in
\hoffset= - 1.0in         

\begin{document}

\vspace{1cm}

\begin{center}

\begin{Large}\fontfamily{cmss}
\fontsize{17pt}{27pt}
\selectfont
	\textbf{Hopf link invariants and integrable hierarchies}
	\end{Large}
	
\bigskip \bigskip
\begin{large}
Chuanzhong Li$^a$ \footnote{lichuanzhong@sdust.edu.cn},
A. Mironov$^{b,c,d}$ \footnote{mironov@lpi.ru; mironov@itep.ru},
A. Yu. Orlov$^{e}$ \footnote{orlovs55@mail.ru}
 \end{large}
 \\

\vspace{1cm}

\begin{small}
$^a$ {\it College of Mathematics and Systems Science, Shandong University of Science and Technology, Qingdao,
266590, China}\\
$^b$ {\it Lebedev Physics Institute, Moscow 119991, Russia}\\
$^c$ {\it ITEP, Moscow 117218, Russia}\\
$^d$ {\it Institute for Information Transmission Problems, Moscow 127994, Russia}\\
$^e${\em Shirshov Institute of Oceanology, RAS, Nahimovskii Prospekt 36, Moscow 117997, Russia }
\end{small}
 \end{center}

\vspace{1cm}

\begin{abstract}
The goal of this note is to study integrable properties of a generating function of the HOMFLY-PT invariants of the Hopf link colored with different representations.
We demonstrate that such a generating function is a $\tau$-function of the KP hierarchy. Furthermore, this Hopf generating function in the case of composite representations, which is a generating function of the 4-point functions in topological string (corresponding to the resolved conifold with branes on the four external legs), is a $\tau$-function of the universal character(UC) hierarchy put on the topological locus. We also briefly discuss a simple matrix model associated with the UC hierarchy.
\end{abstract}


\bigskip

\section{Introduction\label{introduction}}

Integrable properties in a theory are known to allow one to realize group theory structures underlying the theory. In particular, it has been a goal for years to find out such properties in knot theory. Unfortunately, integrability structures in knot theory has been realized so far only for torus knots and links \cite{MMM1}. Moreover, these structures are rather related to braid than to knot invariants: in order to construct a KP $\tau$-function out of invariant, one has to get off the topological locus \cite{MMM1}. In fact, one can reproduce the HOMFLY-PT polynomials of the torus knots and links via a matrix model \cite{knotmamo}, and matrix models are typically $\tau$-functions of integrable hierarchies. But, again, their deformations to non-torus links meets serious problems \cite{femamo}.

On the other hand, integrability structures are often found in string theory \cite{UFN3}. Now we notice that the Hopf HOMFLY-PT polynomial is associated with a brane pattern in topological string theory \cite{tv,Aganagic:2003db,IKV,GIKV,AK}. Hence, one may expect that, in the case of the Hopf link, integrability structures should be especially profound. The goal of the present note to demonstrate that this is, indeed, the case, and the generating function of the Hopf HOMFLY-PT invariants in various representations is immediately a $\tau$-function of the KP hierarchy.

To be precise, we consider the Hopf HOMFLY-PT invariants with two components of the Hopf link colored with representations $\lambda$ and $\mu$ of $SL(N)$, and introduce a generating function of these invariants:
\be\label{Hg}
Z(p,\bar p)=\sum_{\lambda,\mu}S_{\mu}(p)S_\lambda(\bar p)H^{hopf}_{\lambda,\mu}(q,N)
\ee
where $S_{\mu}(p)$ is the Schur function labeled by the Young diagram (or partition)\footnote{Hereafter, we merely say representation $\lambda$ implying the representation associated with the Young diagram $\lambda$.} $\mu$. We will prove that this generating function is a $\tau$-function of the KP hierarchy with respect to both the time variables\footnote{Note that time variables of the KP hierarchy are usually chosen slightly differently: $t_k=p_k/k$. We use this notation as more common for theory of symmetric functions.}  $p_k$ and, $\bar p_k$.

Moreover, one can consider the Hopf link with components colored with composite representations.
The composite representation is the most general finite-dimensional irreducible highest weight representations of $SU(N)$  \cite{Koike,GW,Vafa,Kanno,MarK}, which are associated with the Young diagram obtained by putting $\lambda$ atop of
$\mu_1$ lines of the lengths $N-\mu_i^{\vee}$, i.e.
\be
[\lambda,\mu]= \Big[\lambda_1+\mu_1,\ldots,\lambda_{l_\lambda}+\mu_1,\underbrace{\mu_1,\ldots,\mu_1}_{N-l_{\!_\lambda}-l_{\!_\mu}},
\mu_1-\mu_{_{l_{\!_\mu}}},\mu_1-\mu_{{l_{\!_\mu}-1}},\ldots,\mu_1-\mu_2\Big].
\ee
Throughout the note, we denote $l_\lambda$ the number of lines in the Young diagram $\lambda$, and denote $\lambda^\vee$ the transposed diagram of $\lambda$.

The basic special function associated with representation is its character
expressed through the Schur functions,
$S_{(\lambda,\varnothing)}=S_{\lambda}$.
For composite representations one needs their generalization, the
composite Schur functions \cite{Koike,Kanno,MMHopf}.
Because of explicit $N$ dependence, they are not easy to define for
arbitrary (generic) values of power sums $P_k$.
Fortunately in most applications we need their reductions to
just $N$-dimensional loci:
\be\label{37}
P_k  =\sum_{i=1}^N x_i^k, \ \ \ \ \ \
P_{-k}  =\sum_{i=1}^N x_i^{-k}
\ee
At these peculiar locus, the composite Schur functions
can be defined by the uniformization trick of \cite{MMHopf},
and they possess a nice description as a bilinear combination
of {\it skew} Schur functions. For an arbitrary composite representation
 $(\lambda,\mu)$ made from a pair of Young diagrams $\lambda$ and $\mu$, there is the Koike formula \cite{Koike}
\be\label{Koike1}
S_{[\lambda,\mu]}\{x_i\} =\left(\prod_{i=1}^Nx_i\right)^{l_{\mu^\vee}}
\sum_{\eta  } (-)^{|\eta|}\cdot S_{\lambda/\eta}(P_k)\cdot S_{\mu/\eta^\vee}(P_{-k})
\ee
where $l_{\mu^\vee}$ denotes the number of lines of the conjugated Young diagram  $\mu^\vee$, and $S_{\lambda/\mu}$ denotes the skew Schur function. The pre-factor is absent in the $SU(N)$ case when $\prod_{i=1}^Nx_i=1$.

One can definitely also define a more general polynomial, which depends on two abstract sets of time variables $P$ and $\bar P$:
\be\label{Koike2}
S_{[\lambda,\mu]}(P,\bar P)=\sum_\eta (-1)^{|\eta|}S_{\lambda/\eta}(P)S_{\lambda/\eta^\vee}(\bar P),
\ee {\bf and this is just the so-called universal character(UC)\cite{Koike,Tsuda}.}
What is important, the Hopf HOMFLY-PT polynomial with components colored with composite representations, $H^{hopf}_{[\lambda,\mu],[\lambda',\mu']}(q,N)$ is associated with
four-point function on the resolved conifold geometry given by gluing two topological vertices\footnote{Four representations $\lambda$, $\lambda'$, $\mu$ and $\mu'$ sit on the four external legs.} \cite{AKMMM}. Hence, the generating function of such polynomials gives simultaneously the generating function of the four-point functions. As we will demonstrate, this generating function is a $\tau$-function of an extension of the KP hierarchy, the UC hierarchy \cite{Tsuda} specialized to a particular topological locus. This should not come as a surprise since the UC hierarchy is just an extension of the KP hierarchy to the universal characters \cite{Koike} associated with composite representations.

Hence, the plan of our note is as follows. In section 2, we describe the UC hierarchy and its properties. In particular, we discuss how to make a single UC $\tau$-function out of two KP $\tau$-functions and explain this constriction in the example of a matrix model. In section 3, following \cite{MMM1}, we demonstrate how a KP integrable structure behind the torus knots/links emerges upon shifting off the topological locus. In section 4, we first prove that the generating function (\ref{Hg}) is a KP $\tau$-function, and then extend the result to the generating function of the Hopf HOMFLY-PT polynomials in composite representations. This generating function is a $\tau$-function of the UC hierarchy, however, in this case, one has to get off the topological locus. Section 5 contains some concluding remarks.

\paragraph{Notation.} The Schur function $S_{\mu}\{x\}$ is a symmetric function of variables $x_i$ \cite{Mac}, and we use the notation $S_{\mu}(p)$ for the Schur function treated as a graded polynomial of the power sums $p_k:=\sum_ix_i^k$. The Schur function is labeled by the Young diagram (partition) $\mu$ with $l_\mu$ lines ($l_\mu$ parts): $\mu_1\ge \mu_2\ge\ldots\ge \mu_{l_\mu}>0$, and the grading of this Schur function is $|\mu|:=\sum_i\mu_i$. We also denote through $S_{\lambda/\mu}$ the skew Schur functions. We use the notation $\mu^\vee$ for the transposed Young diagram.

\section{UC hierarchy}

In this section, we describe the UC hierarchy and its basic properties.

\subsection{$\tau$-function of KP hierarchy and decoupling Pl\"ucker relations}

We start with reminding the standard properties of $\tau$-functions of the KP hierarchy. A wide class of solutions to the KP hierarchy is realized by $\tau$-functions that are formal power series in time variables $p_k$. One can always expand such $\tau$-functions into the Schur functions since they form a complete basis:
\be\label{1}
\tau(p)=\sum_\lambda \pi_\lambda S_\lambda(p).
\ee
The $tau$-function (\ref{1}) is a $\tau$-function of the KP hierarchy iff the expansion coefficients $\pi_\lambda$ satisfy the Pl\"ucker relations in the infinite-dimensional Grassmannian \cite{Sato,DJKM}.

Note that, as was explained in \cite{GKM2,OS,AMMN1}, any function of the form $\dP_\lambda=\exp\Big(\sum_ig(\lambda_i)\Big)$, which is equivalent \cite{AMMN2} to\footnote{This product is sometimes called the generalized content product, because $j-i$ is called the content of the box with coordinates $(i,j)$ in the Young diagram of $\lambda$ \cite{Mac}.} $\dP_\lambda=\prod_{\lambda_{i,j}\in\lambda}\dP(j-i)$ trivially decouples from the Pl\"ucker relations, i.e. if $\pi_\lambda$ is a solution the Pl\"ucker relations, then such is $\pi_\lambda\dP_\lambda$. We call such solutions $\dP_\lambda$ decoupling solutions. The corresponding $\tau$-functions are called the hypergeometric $\tau$-functions \cite{GKM2,OS}.

\subsection{$\tau$-function of UC hierarchy}

The UC hierarchy defined in \cite{Tsuda} is solved by a $\tau$-function, depending on two infinite sets of times variables $P_k$ and $\bar P_k$, which can be generally written in the form
\be
\tau^{UC}(P,\bar P;\pi)=\sum_{\lambda,\mu} S_{[\lambda,\mu]}(P,\bar P)\pi_{\lambda,\mu}
\ee
with $S_{[\lambda,\mu]}(P,\bar P)$ from (\ref{Koike2}).
Here $\pi_{\lambda,\mu}$ is an arbitrary solution to the infinite systems of Pl\"ucker relations in the infinite-dimensional Grassmannian \cite{Sato,DJKM} w.r.t. both $\lambda$ and $\mu$. In other words, an arbitrary solution of the UC hierarchy is associated with a point of the product of two Grassmannians.

In this note, we consider a particular $\tau$-function of the UC hierarchy defined as a sum
\be\label{tau}
\tau^{UC}(p,\bar p;P,\bar P;\dP,\bar\dP)=\sum_{\lambda,\mu}S_\lambda(p)\dP_\lambda S_{[\lambda,\mu]}(P,\bar P)\bar\dP_\mu S_{\mu}(\bar p)
\ee
depending on two infinite sets of time variables $p_k$ and $\bar p_k$.  Here $\dP_\lambda$, $\bar\dP_\mu$ are decoupling solutions to the Pl\"ucker relations, i.e. those of the form $\dP_\lambda=\prod_{\lambda_{i,j}}f(j-i)$, where $f(x)$ is an arbitrary function non-singular at integer points. Since the Schur function $S_\lambda$ satisfies the Pl\"ucker relations, so does the product $S_\lambda\dP_\lambda$.

Such $\tau$-functions may emerge both in matrix models, and in knot theory, and they are a UC hierarchy counterpart of the hypergeometric $\tau$-functions \cite{GKM2,OS}.

\subsection{UC $\tau$-function as a KP $\tau$-function}

To begin with, we consider the simplest $\tau$-function
\be\label{tau0}
\tau_0(p,\bar p;P,\bar P)=\sum_{\mu,\lambda}S_\lambda(p)S_{[\lambda,\mu]}(P,\bar P)S_{\mu}(\bar p).
\ee
Then, using the Cauchy identity,
\be
&&\sum_{\xi} (-Q)^{|\xi|} \cdot S_{\xi/\eta_1}(p)\cdot S_{\xi^\vee/\eta_2^\vee}(p')=
\sum_{\xi} (-Q)^{|\xi|} \cdot S_{\xi/\eta_1}(p)\cdot S_{\xi/\eta_2}((-1)^{k+1}p'_k)\nn\\
&=&\exp\left(-\sum_k \frac{Q^kp_kp_k'}{k}\right) \cdot
\sum_\sigma (-Q)^{|\eta_1|+|\eta_2|-|\sigma|}\cdot
 S_{\eta_1^\vee /\sigma}(p')\cdot S_{\eta_2/\sigma^\vee}(p)
\ee
one immediately obtains
\be\label{sum}
\sum_{\mu}S_{[\lambda,\mu]}(P,\bar P)S_{\mu}(\bar p)=\exp\left(\sum_k{\bar p_k\bar P_k\over k}\right)S_\lambda(P_k-\bar p_k)
\ee
and
\be\label{KP-KP-UC}
\tau_0(p,\bar p;P,\bar P)=\exp\left(\sum_k{\bar p_k\bar P_k\over k}\right)\sum_\lambda S_\lambda(p_k)S_\lambda(P_k-\bar p_k)=
\exp\left(\sum_k{p_kP_k+\bar p_k\bar P_k-p_k\bar p_k\over k}\right)
\ee
which is a trivial KP $\tau$-function w.r.t. the both sets of time variables $p_k$ and $\bar p_k$. In fact, in order to check that (\ref{tau0}) is a KP $\tau$-function w.r.t. to $p_k$'s, it is sufficient to check that the sum (\ref{sum}) satisfies the Pl\"ucker relations, which actually does because the Schur polynomial satisfies them. Similarly, if one chooses $\dP_\mu=\prod_{\mu_{i,j}\in\mu}f(j-i)$ with an arbitrary function $f(x)$ (keeping $\bar\dP_\mu$ trivial), this does not influence the Pl\"ucker relations: such $\dP_\mu$ decouples from the Pl\"ucker relations. Hence, in this case, (\ref{tau}) is still a KP $\tau$-function w.r.t. $p_k$'s but not to $\bar p_k$'s. One can definitely make a similar KP $\tau$-function w.r.t. $\bar p_k$'s choosing $\bar\dP_\mu=\prod_{\mu_{i,j}\in\mu}f(j-i)$ and keeping $\bar\dP_\mu$ trivial.

\subsection{The UC $\tau$-function from two Hermitian Gaussian matrix models}

Similarly to the KP hierarchy, some of the $\tau$-functions of the UC hierarchy can be associated with partition functions of matrix models. In fact, since the UC hierarchy is constructed from two KP hierarchies, one expects that some $\tau$-functions of the UC hierarchy can be made from two Hermitian matrix models, each of them being a $\tau$-function of the KP hierarchy. Here we describe a manifest way to glue two matrix model KP $\tau$-functions to a single UC $\tau$-function.

To this end, consider the matrix model partition function
\be\label{makp}
Z(p,\bar p)=\int dU Z^+(p,U)Z^-(\bar p,U^{-1})
\ee
where
\be
Z^\pm(p)=\int dH \exp\left(-{1\over 2}\Tr H^2+\sum_k{p_k\pm\Tr U^k\over k}\Tr H^k\right)
\ee
where $U$ and $H$ are the $N\times N$ unitary and Hermitian matrices accordingly, $dU$ being the Haar measure on the unitary matrices, and $dH$, on the Hermitian ones. $Z^\pm(p)$ is a $\tau$-function of the Toda chain hierarchy \cite{GMMMO,KMMOZ}, and, hence, of the KP hierarchy too.

Let us note that the partition function of the Hermitian matrix model with the Gaussian potential can be presented in the form \cite{SI}
\be\label{15}
Z^\pm(p)=\sum_\lambda \eta_\lambda(N)S_\lambda(p_k=\delta_{k,2})S_\lambda\left(p_k\pm\Tr U^k\right)=
\sum_\lambda \eta_\lambda(N)S_\lambda(p_k=\delta_{k,2})S_{\lambda/\nu}(p_k)S_\nu(\pm\Tr U^k)
\ee
where $\eta_\lambda(N)=\prod_{\lambda_{i,j}\in\lambda}(N-j+i)$ is the content product.
Now, using that
\be
S_\nu(-\Tr U^k)=(-1)^{|\nu|}S_{\nu^\vee}(\Tr U^k),\nn\\
\int dUS_\mu(\Tr U^k)S_\nu(\Tr U^k)=\delta_{\mu\nu},
\ee
one immediately obtains for $Z(p,\bar p)$:
\be
Z(p,\bar p)=\sum_{\lambda,\mu,\nu}(-1)^{|\nu|} \eta_\lambda(N)S_\lambda(p_k=\delta_{k,2})S_{\lambda/\nu}(p_k)
\eta_\mu(N)S_\mu(p_k=\delta_{k,2})S_{\mu/\nu^\vee}(\bar p_k)\nn\\
=\sum_{\lambda,\mu}\pi_\lambda(N)S_{[\lambda,\mu]}(p_k,\bar p_k)
\pi_\mu(N)
\ee
where $\pi_\lambda(N)=\eta_\lambda(N)S_\lambda(p_k=\delta_{k,2})$ is a solution to the Pl\"ucker relations since $\eta_\lambda(N)$ is a decoupling solution. Thus, indeed, $Z(p,\bar p)$ is a $\tau$-function of the UC hierarchy. More complicated examples of matrix models associated with UC $\tau$-functions (of type \cite{NO-Dubrovin,Belowezhie,AntOrlVas}) are discussed elsewhere \cite{LMO1}.

\subsection{Constructing UC $\tau$-function from two KP $\tau$-functions}

One can apply this construction to glue arbitrary two KP $\tau$-functions $\tau^{KP}_{1,2}(p_k)$ into a single UC $\tau$-function $\tau^{UC}(p_k,\bar p_k)$. The simplest idea is just to repeat the trick with the unitary matrix integration:
\be
\tau^{UC}(p_k,\bar p_k)=\int dU\tau^{KP}_{1}(p_k+\Tr U^k)\tau^{KP}_{2}(\bar p_k-\Tr U^{-k}).
\ee
There is, however, a subtlety here: in expansion (\ref{15}), only partitions with the number of parts $l_\nu$ not more than $N$ contributes, $l_\nu\le N$. Since the KP $\tau$-function (\ref{makp}) automatically guarantees this because of the factor $\eta(N)$, it does not give rise to restrictions. However, in the case of generic KP $\tau$-functions this would lead to a new restriction. In order to avoid this restriction, one has to consider $N\to\infty$, which requires some care.

There is another possibility: one can realize $\tau^{UC}(p_k,\bar p_k)$ as
\be\label{uckp}
\tau^{UC}(p_k,\bar p_k)=\left.\left\langle\tau^{KP}_{1}(p_k+P_k)\right|\tau^{KP}_{2}(\bar p_k-P_k)\right\rangle
\ee
where, for the parametrization $P_\Delta=\prod_{k=1} P_k^{m_k}$, the scalar product is defined to be \cite{Mac}
\be
\langle P_\Delta|P_{\Delta'}\rangle=z_\Delta\delta_{\Delta,\Delta'}
\ee
and $z_\Delta:=\prod_k k^{m_k}m_k!$ is the standard symmetric factor of the Young diagram (order of the automorphism). The crucial property of the scalar product that allows one to repeat the above derivation of the UC $\tau$-function is that
\be
\langle S_\lambda|S_\mu\rangle=\delta_{\lambda\mu}.
\ee

The scalar product (\ref{uckp}) can be also realized manifestly (for polynomials $f_1$, $f_2$)
\be
\langle f_1(P_k)|f_2(P_k)\rangle=\left.f_1\left({1\over k}{\p\over\p P_k}\right)f_2(P_k)\right|_{P_k=0}
\ee
which gives rise to formula \cite{Tsuda}:
\be
\tau^{UC}(p_k,\bar p_k)=\left.\tau^{KP}_{1}\left(p_k+{1\over k}{\p\over\p P_k}\right)\tau^{KP}_{2}(\bar p_k-P_k)\right|_{P_k=0}.
\ee

\section{Getting off the topological locus: $\tau$-functions and torus knots/links}


Let us now discuss the $\tau$-function that naturally emerges as a generating function of knot invariants for the torus knots and links. We follow here the construction of \cite{MMM1}.

The unreduced HOMFLY-PT polynomial for the torus knot/link is given by the celebrated Rosso-Jones formula \cite{RJ,China}. Consider torus link with $r$ components and co-prime $n$ and $m$: $T^{[nr,mr]}$ ($r=1$ corresponds to the torus knot) colored with the representations associated with the Young diagrams $\lambda^{(1)},\ldots,\lambda^{(r)}$ of $A_{N-1}$. Then, the HOMFLY-PT polynomial is given by formula
\be\label{HOMFLY}
H^{(n,m;r)}_{\lambda^{(1)},\ldots,\lambda^{(r)}}=q^{mn\sum_i\varkappa_{\lambda^{(i)}}}\sum_{\mu\vdash n\sum_i|\lambda^{(i)}|} q^{-{m\over n}\varkappa_\mu}c_{\lambda^{(1)},\ldots,\lambda^{(r)}}^\mu S_\mu(p^\ast_k).
\ee
Here $\varkappa_\lambda$ is the eigenvalue of the second Casimir operator in representation $\lambda$,
\be\label{Cas}
\varkappa_\lambda=2\kappa_\lambda-{|\lambda|^2\over N}+|\lambda|N
\ee
with $\kappa_\lambda=\sum_{\lambda_{i,j}\in \lambda}(j-i)$, where the sum goes over the boxes of the Young diagram $\lambda$ and $\kappa_\lambda$ is
the corresponding eigenvalue of the cut-and-join operator \cite{MMN},
\be\label{cj}
\hat W_2 S_\lambda &=& \kappa_\lambda S_\lambda,\nn\\
\hat W_2:&=&{1\over 2}\sum_{a,b\ge 1}\left((a+b)p_ap_b{\p\over\p p_{a+b}}+abp_{a+b}{\p^2\over\p p_a\p p_b}\right)
\ee
the Schur polynomial is taken at
\be\label{toploc}
p^\ast_k:={q^{Nk}-q^{-Nk}\over q^k-q^{-k}}
\ee
called the topological locus, and $c_{\lambda_1,\ldots,\lambda_r}^\mu $ are the Adams coefficients defined from
\be\label{Adams}
\prod_iS_{\lambda^{(i)}}(p_{nk})=\sum_{\mu\vdash n\sum_i|\lambda^{(i)}|}c_{\lambda^{(1)},\ldots,\lambda^{(r)}}^\mu S_\mu(p_k).
\ee

Formula (\ref{HOMFLY}) is in the topological framing, and is associated with realization of the torus link by the closed $n$-strand braid with repeated patters within the braid \cite{MMM1}. In fact, the same invariant admits many other braid realization, in particular, $n$ and $m$ may be interchanged, the result for the HOMFLY-PT being the same, which is guaranteed by the condition of topological locus (\ref{toploc}) and by choosing the topological framing.

In \cite{MMM1}, we explained that this expression can be naturally extended from the topological locus (\ref{toploc}) to arbitrary variables $p_k$:
 \be\label{braid}
P^{(n,m;r)}_{\lambda^{(1)},\ldots,\lambda^{(r)}}(p)=\sum_{\mu\vdash n\sum_i|\lambda^{(i)}|} q^{-{2m\over n}\kappa_\mu}c_{\lambda^{(1)},\ldots,\lambda^{(r)}}^\mu S_\mu(p_k).
\ee
Here we changed the framing factor for a more convenient one, since the topological framing is no longer distinguished: the polynomial at arbitrary $p_k$ is no longer a topological invariant anyway.

Though the polynomial (\ref{braid}) is no longer a knot/link invariant, it is associated with the closed $n$-strand braid. Let us consider a generating function of such polynomials:
\be
Z^{(n,m;r)}(p;g^{(1)},\ldots,g^{(r)}):=
\sum_{\lambda^{(1)},\ldots,\lambda^{(r)}} P^{(n,m;r)}_{\lambda^{(1)},\ldots,\lambda^{(r)}}(p)\prod_iS_{\lambda^{(i)}}(g^{(i)}_{k}).
\ee
Now, one can rewrite (\ref{braid})
\be
P^{(n,m;r)}_{\lambda^{(1)},\ldots,\lambda^{(r)}}(p)\stackrel{(\ref{cj})}{=}q^{-{2m\over n}\hat W_2}\sum_{\mu\vdash n\sum_i|\lambda^{(i)}|} c_{\lambda^{(1)},\ldots,\lambda^{(r)}}^\mu S_\mu(p_k)\stackrel{(\ref{Adams})}{=}q^{-{2m\over n}\hat W_2}\cdot
\prod_iS_{\lambda^{(i)}}(p_{nk})
\ee
and
\be\label{tltau}
Z^{(n,m;r)}(p;g^{(1)},\ldots,g^{(r)})\stackrel{Cauchy\ identity}{=}
q^{-{2m\over n}\hat W_2}\cdot \underbrace{\exp\left(\sum {\sum_i g^{(i)}_k\over k}\cdot p_{nk}\right)}_{\tau_b}.
\ee
The generating function of topological invariants is obtained from this function at the topological locus: $Z^{(n,m;r)}(p^\ast;g^{(1)},\ldots,g^{(r)})$.

Note that $\tau_b$ is a trivial KP $\tau$-function being a linear exponential in time variables, and the cut-and-join operator is an element of the $W_{1+\infty}$-algebra \cite{MMMP1}. Hence, following \cite{Orlov,TT}, we come to the conclusion that {\bf $Z^{(n,m;r)}(p;g^{(1)},\ldots,g^{(r)})$ is a KP $\tau$-function in variables $p_k$'s}.

\section{$\tau$-functions and Hopf link}

\subsection{From the Hopf link to a $\tau$-function}

In this section, we deal with the simplest two-component link: the Hopf link. Its HOMFLY-PT polynomial with two components colored with representations $\lambda$ and $\mu$ is
\be\label{Hopf}
H^{hopf}_{\lambda,\mu}(q,N)=S_{\lambda}(p^\ast)\cdot S_{\mu}\Big(P^{(\lambda)}(q,N)\Big)
\ee
where
\be\label{Ph}
P^{(\lambda)}_k(q,N):=
 q^{{2|\lambda|\over |N|}k}\left(p_k^\ast+ q^{-Nk}\cdot   \sum_{j=1}^{l_{\!_\lambda}}  q^{(2j-1)k}\cdot(q^{-2k\lambda_j}-1)
\right).
\ee
Formula (\ref{Hopf}) is symmetric w.r.t. permuting components, i.e. $\lambda\leftrightarrow \mu$. This polynomial is given in the canonical framing \cite{CFA,MarCF} (see a detailed discussion of this issue in \cite{Bai}), and, for the sake of simplicity, from now on, we drop the factor $ q^{{2|\lambda|\over |N|}k}$ in (\ref{Ph}), which is in charge of difference between $U(N)$ and $SU(N)$ invariants.

Now one can consider a generating function w.r.t. to one of the components:
\be
Z_{\lambda}(p):&=&\sum_{\mu}S_{\mu}(p)H^{hopf}_{\lambda,\mu}(q,N)=
S_{\lambda}(p^\ast)
\exp\left(\sum_k{P^{(\lambda)}(q,N)\over k}p_k\right).
\ee
As an exponential linear in time variables $p_k$, it is a trivial KP $\tau$-function.

Let us note that $Z_{\lambda}(p)$ satisfies the Pl\"ucker relations since the exponential as a function of the Young diagram $\lambda$ is a decoupling solution to the Pl\"ucker relations.

Hence, one can go further and convolve $Z_{\lambda}(p)$ with the Schur polynomial to obtain a full generating function:
\be
Z(p,\bar p)=\sum_\lambda Z_{\lambda}(p)S_\lambda(\bar p)=\sum_{\lambda,\mu}S_{\mu}(p)S_\lambda(\bar p)H^{hopf}_{\lambda,\mu}(q,N).
\ee
Since $Z_{\lambda}(p)$ satisfies the Pl\"ucker relations, {\bf $Z(p,\bar p)$ is a KP $\tau$-function in time variables $\bar p_k$}. However, as a corollary of symmetry of (\ref{Hopf}) in $\lambda$ and $\mu$, {\bf $Z(p,\bar p)$ is also a KP $\tau$-function in time variables $p_k$}.

In fact, there is another way to see this integrability: one can use a matrix model representation of the Hopf HOMFLY-PT polynomial \cite{knotmamo,femamo}\footnote{In terms of \cite[Eq.(2)]{femamo}, we use the variables $h_i=e^{u_i}$ and then treat it as an eigenvalue matrix integral in $h_i$ coming from the Hermitian matrix integral.}:
\be
H^{hopf}_{\lambda,\mu}(q,N)\sim\int dH\exp\left(-{1\over 4\ln q}\Tr \ln^2H-N\Tr\ln H\right)S_\lambda(H)S_\mu(H)
\ee
where $H$ are $N\times N$ Hermitian matrices. Indeed, this representation implies that
\be
Z_N(p,\bar p)=\int dH\exp\left(-{1\over 4\ln q}\Tr \ln^2H-N\Tr\ln H+\sum_k{\Tr H^k\over k}(p_k+\bar p_k)\right)
\ee
and this is a KP $\tau$-function\footnote{However, this is not a $\tau$-function of the Toda chain where $N$ is the zeroth time variable because of an explicit dependence of the matrix model potential on $N$.}  \cite{GMMMO,KMMOZ,UFN3} w.r.t. the both sets of times $p_k$ and $\bar p_k$ and, moreover, depends on the sum $p_k+\bar p_k$, as it should be for the Toda chain. 

This matrix model expression for the generating function can be naturally uplifted to the $q,t$-deformed case  \cite{MMP}. In this case, the Hopf link HOMFLY-PT polynomial is substituted by the Hopf hyperpolynomial convolved with the Macdonald polynomials instead of the Schur functions, and the generating function is an $N$-fold eigenvalue integral. This integral still depends on the sum of times $p_k$ and $\bar p_k$, though integrability under the $q,t$-deformation is as usual lost (because of the deformation of the square of the Vandermonde determinant, which provided integrable determinant representation of the eigenvalue integral in the non-deformed case).

\subsection{Hopf link in composite representations}

Now we consider the Hopf link HOMFLY-PT polynomial with components colored with composite representations \cite{MMHopf}.
This HOMFLY-PT polynomial is associated with a topological string object, that is, with the resolved conifold with four different representations on the four external legs, and can be realized as a four-point function made of two topological vertices \cite{AKMMM}.
This HOMFLY-PT polynomial has the form \cite{MMHopf,AKMMM}
\be\label{Hopfc}
H^{hopf}_{[\lambda,\mu],[\lambda'\mu']}(q,N)=S_{[\lambda,\mu]}(p^\ast,p^\ast)\cdot S_{[\lambda',\mu']}\Big(P^{(\lambda,\mu)}(q),
P^{(\lambda,\mu)}(q^{-1})\Big)
\ee
where
\be\label{P}
P^{(\lambda,\mu)}(q):=
 q^{2{|\lambda|-|\mu|\over N}k}\left(p_k^\ast+ q^{-Nk}\cdot   \sum_{j=1}^{l_{\!_\lambda}}  q^{(2j-1)k}\cdot(q^{-2k\lambda_j}-1) + q^{Nk}\cdot  \sum_{i=1}^{l_\mu}q^{(1-2i)k}\cdot(q^{ 2k\mu_i}-1)
\right).
\ee
Formula (\ref{Hopfc}) is symmetric w.r.t. permuting components, i.e. $(\lambda,\mu)\leftrightarrow (\lambda',\mu')$. This polynomial is given in the canonical framing \cite{CFA,MarCF,Bai}, and, for the sake of simplicity, we again drop the factor $ q^{2{|\lambda|-|\mu|\over N}k}$ in (\ref{P}).

Now one can consider a generating function w.r.t. to one of the components:
\be
Z_{[\lambda,\mu]}(g^{(1)},g^{(2)}):&=&\sum_{\lambda',\mu'}
S_{\lambda'}(g^{(1)})H^{hopf}_{[\lambda,\mu],[\lambda'\mu']}(q,N)S_{\mu'}(g^{(2)})=
S_{[\lambda,\mu]}(p^\ast,p^\ast) \tau_0(g^{(1)},g^{(2)};P^{(\lambda,\mu)}(q),
P^{(\lambda,\mu)}(q^{-1})\nn\\
&=&S_{[\lambda,\mu]}(p^\ast,p^\ast)
\exp\left(\sum_k{g^{(1)}_kP^{(\lambda,\mu)}_k(q)+g^{(2)}_k P_k^{(\lambda,\mu)}(q^{-1})-g^{(1)}_kg^{(2)}_k\over k}\right).
\ee
As we explained above, this is a trivial KP $\tau$-function w.r.t. variables $g^{(1)}$ and $g^{(2)}$.

Now let us convolve this partition function with two more Schur functions in order to obtain the general generating function:
\be
Z(g^{(1)},g^{(2)},g^{(3)},g^{(4)}):&=&\sum_{\lambda,\mu}Z_{[\lambda,\mu]}(g^{(1)},g^{(2)})S_\lambda(g^{(3)})S_\mu(g^{(4)}) \\
&=&
\exp\left(\sum_k{g^{(1)}_kp^\ast_k+g^{(2)}_kp^\ast_k-g^{(1)}_kg^{(2)}_k\over k}\right)
\tau^{UC}\Big(g^{(3)},g^{(4)};p^\ast,p^\ast;\dP(g^{(1)},g^{(2)};q),\dP(g^{(1)},g^{(2)};q^{-1})\Big),\nn
\ee
with
\be
\dP_\lambda(p',p'';q)=\exp\left(\sum_k{\phi_{\lambda,k}(q)p'_k+\phi_{\lambda,k}(q^{-1})p''_k\over k}\right),\nn\\
\phi_{\lambda,k}(q):=q^{-Nk}\sum_{j=1}^{l_{\!_\lambda}}  q^{(2j-1)k}\cdot(q^{-2k\lambda_j}-1).
\ee
Such $\dP_\lambda$, as we explained above, satisfies the Pl\"ucker relations. Hence, if one extends this expression off the topological locus $p_k^\ast$ to arbitrary time variables $P_k$ and $\bar P_k$:
\be
Z(P,\bar P)=
\exp\left(\sum_k{g^{(1)}_kp^\ast_k+g^{(2)}_kp^\ast_k-g^{(1)}_kg^{(2)}_k\over k}\right)
\tau^{UC}\Big(g^{(3)},g^{(4)};P,\bar P;\dP(g^{(1)},g^{(2)};q),\dP(g^{(1)},g^{(2)};q^{-1})\Big),
\ee
and {\bf  the generating function $Z(P,\bar P)$ is a UC $\tau$-function of the form (\ref{tau}) in variables $P$, $\bar P$}, i.e.
\be
Z(P_k=p^\ast_k,\bar P_k=p_k^\ast)=Z(g^{(1)},g^{(2)},g^{(3)},g^{(4)}).
\ee

\section{Conclusion}

In this note we demonstrated that a generating function of the HOMFLY-PT polynomials of the Hopf link with components colored with different representations is a $\tau$-function of the KP hierarchy. In variance with previously known integrable structures in torus knots/links \cite{MMM1}, this claim does not require getting off the topological locus.

The generalization of this claim to the composite representations in the Hopf link components already requires getting off the topological locus so that the proper generating function of the Hopf link is obtained from the $\tau$-function of the UC hierarchy only at a peculiar point in the space of time variables. In this sense, the composite representation case is a bit similar to all other torus knots/links situation. However, we notice that the Hopf link with components colored with composite representations $[\lambda,\mu]$ and $[\lambda',\mu']$ is equivalent \cite{AKMMM} to a special (in a sense, maximal) projection of the colored HOMFLY-PT invariant ${\cal H}^{L_{8n8}}$ of the link $L_{8n8}$ (in accordance with the Thistlethwaite link table \cite{twi}) in the colored space. Since the link $L_{8n8}$ is not torus, we obtain the first example of an integrability structure in the non-torus case.

The described structures can be also reformulated in matrix model terms. We discuss this issue as well as the issue of other relations of matrix models with the UC hierarchy elsewhere \cite{LMO1}.

\section*{Acknowledgements}

Our work is partly supported by grant RFBR 21-51-46010 ST-a (A.Mir.), by the grant of the Foundation for the Advancement of Theoretical Physics and Mathematics ``BASIS" (A.Mir.) and by the National Natural Science Foundation of China under Grants No. 12071237.


\begin{thebibliography}{99}

\bibitem{MMM1}  A. Mironov, A. Morozov, An. Morozov,
in: {\it Strings, Gauge Fields, and the Geometry Behind: The Legacy of Maximilian Kreuzer},
eds: A.Rebhan, L.Katzarkov, J.Knapp, R.Rashkov, E.Scheidegger,
World Scientific,   2013 pp.101-118, 
arXiv:1112.5754

\bibitem{knotmamo} M. Tierz, Mod. Phys. Lett. {\bf A19} (2004) 1365-1378, hep-th/0212128\\
A. Brini, B. Eynard, M. Marino, Annales Henri Poincare, {\bf 13} (2012) No.8,  arXiv:1105.2012

\bibitem{femamo} A. Alexandrov, A. Mironov, A. Morozov, An. Morozov,
JETP Letters, {\bf 100} (2014) 271-278, arXiv:1407.3754

\bibitem{UFN3} A. Morozov,
Phys. Usp.(UFN) {\bf 37} (1994) 1;
hep-th/9502091; hep-th/0502010\\
A. Mironov, Int. J. Mod. Phys. {\bf A9} (1994) 4355; Phys. Part. Nucl.
{\bf 33} (2002) 537; hep-th/9409190

\bibitem{tv} M. Aganagic, M. Mari\~no, C. Vafa, 
Commun. Math. Phys. {\bf 247} (2004) 467-512,
hep-th/0206164

\bibitem{Aganagic:2003db}
  M.~Aganagic, A.~Klemm, M.~Marino, C.~Vafa,
  Commun.\ Math.\ Phys.\  {\bf 254} (2005) 425, hep-th/0305132

\bibitem{IKV} A. Iqbal, C. Kozcaz, C. Vafa, 
	JHEP {\bf 0910} (2009) 069, hep-th/0701156

\bibitem{GIKV} S. Gukov, A. Iqbal, C. Kozcaz, C. Vafa, Commun. Math. Phys. {\bf 298} (2010) 757-785, arXiv:0705.1368

\bibitem{AK} H. Awata, H. Kanno,
J. Phys. {\bf A44} (2011) 375201, arXiv:0910.0083

\bibitem{Koike} K. Koike, 
Adv. Math. {\bf 74} (1989) 57

\bibitem{Kanno} H. Kanno, Nucl.Phys. {\bf B745} (2006) 165-175, hep-th/0602179

\bibitem{GW} D. J. Gross, W. Taylor, 
Nucl. Phys. {\bf B400} (1993) 181, hep-th/9301068

\bibitem{Vafa} M. Aganagic, H. Ooguri, N. Saulina, C. Vafa,
Nucl.\ Phys. {\bf B715}  (2005) 304, hep-th/0411280 \\
M. Aganagic, A. Neitzke, C. Vafa, Adv. Theor. Math. Phys. {\bf 10} (2006) 603-656, hep-th/0504054

\bibitem{MarK} M. Mari\~no, Commun. Math. Phys. {\bf 298} (2010) 613-643, arXiv:0904.1088

\bibitem{MMHopf} A. Mironov and A. Morozov, JETP Lett. {\bf 107} (2018)  728-735,   arXiv:1804.10231

\bibitem{AKMMM} H. Awata, H. Kanno, A. Mironov, A. Morozov, An. Morozov,
Phys. Rev. {\bf D98} (2018) 046018,  arXiv:1806.01146

\bibitem{Tsuda} T. Tsuda, 
Commun.Math.Phys. {\bf 248} (2004) 501-526

\bibitem{Mac} I.G. Macdonald,
  \textit{Symmetric functions and Hall polynomials},
  Oxford University Press, 1995

\bibitem{Sato} M.~Sato, 
RIMS Kokyuroku, Kyoto Univ. {\bf 439}  (1981) 30-46

\bibitem{DJKM} E. Date, M. Jimbo, M. Kashiwara, T. Miwa,
in:  {\sl Nonlinear integrable systems-classical theory and quantum theory},
eds. M. Jimbo, and T. Miwa, World Scientific, 39-120, (1983).

\bibitem{GKM2} S. Kharchev, A. Marshakov, A. Mironov, A. Morozov, Int. J. Mod. Phys. {\bf A10} (1995) 2015, hep-th/9312210

\bibitem{OS} A. Orlov, D. M. Shcherbin,
Theor. Math. Phys.
{\bf 128} (2001) 906-926

\bibitem{AMMN1} A.~Alexandrov, A.~Mironov, A.~Morozov, S.~Natanzon,
J. Phys. A \textbf{45} (2012) 045209,
arXiv:1103.4100

\bibitem{AMMN2} A.~Alexandrov, A.~Mironov, A.~Morozov, S.~Natanzon,
JHEP \textbf{11} (2014) 080,
arXiv:1405.1395



\bibitem{GMMMO} A. Gerasimov, A. Marshakov, A .Mironov, A. Morozov, A. Orlov,
Nucl. Phys. {\bf B357} (1991) 565

\bibitem{KMMOZ} S. Kharchev, A. Marshakov, A. Mironov, A. Orlov, A. Zabrodin,
Nucl. Phys. {\bf B366} (1991) 569-601

\bibitem{MMP} A.~Mironov, A.~Morozov and A.~Popolitov,
arXiv:2410.03175

\bibitem{SI} A.~Mironov, A.~Morozov,
  Phys.Lett.\ {\bf B771} (2017) 503,
arXiv:1705.00976

\bibitem{NO-Dubrovin}
S. M. Natanzon, A. Yu. Orlov, 
Proceedings of Symposia in Pure Mathematics, {\bf 103} (2021) 337-377, arXiv:2002.00466

\bibitem{Belowezhie} A.Yu. Orlov, 
in: {\sl Geometric Methods in Physics XXXVIII}: Workshop, Bialowieza, Poland, 2019, pp. 221-250,
Springer International Publishing (2020)

\bibitem{AntOrlVas} E. N. Antonov, A. Yu. Orlov, D. V. Vasiliev,
``Coupling of different solvable ensembles of random matrices'',
arXiv:2310.03732

\bibitem{LMO1} C. Li, A. Mironov, A. Orlov,
``Matrix model solutions of the UC hierarchy of integrable equations'', to appear

\bibitem{RJ} M. Rosso, V.F.R. Jones, J.Knot Theory Ramifications, {\bf 2} (1993) 97-112

\bibitem{China} X. S. Lin, H. Zheng, Trans. Amer. Math. Soc. {\bf 362} (2010) 1-18, math/0601267

\bibitem{MMN} A.~Mironov, A.~Morozov, S.~Natanzon,
Theor. Math. Phys. \textbf{166} (2011) 1-22,
arXiv:0904.4227;
J. Geom. Phys. \textbf{62} (2012) 148-155,
arXiv:1012.0433

\bibitem{MMMP1}  A.~Mironov, V.~Mishnyakov, A.~Morozov, A.~Popolitov,
JHEP \textbf{23} (2020) 065,
arXiv:2306.06623

\bibitem{Orlov} A. Orlov, Plasma theory and nonlinear and turbulent processes in physics, {\bf 1} (1988) 13-25\\
P. Winternitz, A.Yu. Orlov, 
Theoret. i Matem. Fizika (in Russian), {\bf 113} (1997) 231-260 (Theor. Math. Phys. {\bf 113} (1997) 1393-1417)

\bibitem{TT} K.~Takasaki, T.~Takebe, hep-th/9207081;
Lett. Math. Phys. \textbf{28} (1993) 165-176,
hep-th/9301070

\bibitem{CFA} M. Atiyah, 
Topology {\bf 29} (1990) 1

\bibitem{MarCF} M. Mari\~no, C. Vafa, Contemp. Math. {\bf 310} (2002) 185-204, hep-th/0108064\\
M. Mari\~no, {\sl Enumerative geometry and knot invariants}, in: {\sl 70th Meeting between
Physicists, Theorist and Mathematicians}, Strasbourg, France, May 23-25, 2002, hep-th/0210145

\bibitem{Bai} C.~Bai, J.~Jiang, J.~Liang, A.~Mironov, A.~Morozov, A.~Morozov, A.~Sleptsov,
Phys. Lett. {\bf B778} (2018) 197-206,
arXiv:1709.09228

\bibitem{twi} http://katlas.org/wiki/The\_Thistlethwaite\_Link\_Table

\end{thebibliography}
\end{document}